\documentclass[10pt,twocolumn]{article}

\usepackage[margin=1in, columnsep=0.2in]{geometry}
\usepackage{multicol}
\usepackage[font=small,labelfont=bf,justification=justified,width=\textwidth,skip=8pt]{caption}
\usepackage{placeins}
\usepackage{afterpage}
\usepackage{balance}  
\usepackage{dblfloatfix}  
\usepackage{flushend}  
\usepackage{graphicx}
\usepackage{amsmath}
\usepackage{amsfonts}
\usepackage{amssymb}
\usepackage{bm}
\usepackage{units}
\usepackage{yfonts}
\usepackage{subfigure}
\usepackage[table]{xcolor}
\usepackage{soul}
\usepackage{hyperref}
\usepackage{svg}
\usepackage{float}
\usepackage[normalem]{ulem}
\usepackage{authblk}
\usepackage{setspace}
\usepackage[scaled]{helvet}
\usepackage{braket}
\usepackage{multirow}
\usepackage{mathrsfs}
\usepackage{booktabs}
\usepackage{algorithm}
\usepackage{algorithmicx}
\usepackage{algpseudocode}
\usepackage{listings}
\usepackage{xcolor}
\usepackage[utf8]{inputenc}
\usepackage{textcomp}
\usepackage{newunicodechar}
\usepackage{soul}

\newunicodechar{−}{-}

\newcommand{\fighel}{\sffamily}

\newcommand{\red}[1]{{\color{red}#1}}
\newcommand{\green}[1]{{\color{green}#1}}

\title{Experimental Milestones Towards Majorana Braiding with Acoustic Metamaterials}

\author{Jackson Saunders\textsuperscript{1}, Emil Prodan\textsuperscript{2}, Camelia Prodan\textsuperscript{1}}

\date{} 

\renewenvironment{abstract}%
{\small\begin{center}\textbf{Abstract}\end{center}\begin{quotation}\small}%
{\end{quotation}}

\usepackage{titlesec}
\titleformat{\section}{\large\bfseries}{\thesection}{1em}{}
\titleformat{\subsection}{\normalsize\bfseries}{\thesubsection}{1em}{}
\titlespacing*{\section}{0pt}{12pt plus 4pt minus 2pt}{6pt plus 2pt minus 2pt}
\titlespacing*{\subsection}{0pt}{8pt plus 2pt minus 2pt}{4pt plus 2pt minus 2pt}

\begin{document}

\twocolumn[
\begin{@twocolumnfalse}
\maketitle

\begin{center}
\small
\textsuperscript{1} Department of Physics and Engineering Physics, Fordham University, Bronx, NY 10458, USA.\\
\textsuperscript{2} Department of Physics, Yeshiva University, New York, NY 10016, USA.\\

\bigskip
\textsuperscript{*}Correspondence: \href{mailto:Jsaunders26@fordham.edu}{Jsaunders26@fordham.edu, prodan@yu.edu, cprodan@fordham.edu}
\end{center}

\begin{abstract}
Here we show the first experimental implementation of the fully general Kitaev chain with complex-valued order parameter $\Delta$ and site-varying synthetic chemical potential $\mu$, using a passive multilayer acoustic resonator design and fabrication. Our laboratory model faithfully reproduces the key symmetries and the topological phase diagram of the model, and displays robust Majorana-like edge modes spatially localized at smoothly-engineered domain walls and energetically localized in the middle of the bulk spectral gap. We demonstrate precise control over mode positioning through smooth spatial variations of $\mu$, and validate the stability of the modes and of the spectral gap under continuous and complex variations of $\Delta$ — both critical requirements for topological braiding operations. These results establish and validate the fundamental building blocks for experimental implementation of complete braiding protocols, opening concrete pathways toward accessible non-abelian physics and topologically protected information processing.
\end{abstract}


\vspace{0.5cm}
\end{@twocolumnfalse}]

\section{Introduction}

Under specialized designs, the dynamics of classical degrees of freedom of architected materials can emulate a variety of quantum effects \cite{CHEN20251347}, such as electron's half spin \cite{PhysRevLett.130.026101}, the dispersionless propagation of the majorana particles \cite{WU2024893}, or Thouless' topological pumping \cite{WentingPhysRevLett.125.224301}. Metamaterials can be endowed with exact fundamental symmetries such as time-reversal, particle-hole and chiral symmetries, both of  bosonic and fermionic type \cite{PhysRevB.98.094310}. Perhaps the most exciting opportunity opened by these bridges between classical and quantum physics is the access to non-abelian phenomena \cite{doi:10.1126/science.adf9621}. The latter refers to the emergence of a Wilczeck-Zee non-abelian connection on a degenerate resonant level under adiabatic parametric deformations  \cite{PhysRevLett.52.2111}. With enough control, one can implement unconventional methods of communication and information processing that are believed to be more robust against noise and imperfections \cite{Freedman2003}. 

Non-abelian adiabatic braiding of modes has been recently demonstrated with acoustics \cite{Chen2022} and optics \cite{chen2025highdimensional}, though in a non-topological setting. Braiding of degenerate mid-gap states in a coupled-mode theoretic setting has been predicted and numerically demonstrated to be topological \cite{emilbraiding}. The model used in \cite{emilbraiding} was the ``de-complexified" version from Eq.~\ref{Eq:KitaevReal} of the Kitaev chain over a 1-dimensional lattice, which in the original form \cite{Kitaev2001} reads
\begin{equation}
\label{Eq:HamKitaev}
\begin{aligned}
 & H= \tfrac{\imath}{2}( \Delta_{x} \sigma_1+ \Delta_y \sigma_3) \otimes (S - S^\dagger) \\
 & \qquad \qquad \qquad \qquad - \sigma_2 \otimes \big (\mu - \tfrac{t}{2}(S + S^\dagger) \big ).
 \end{aligned}
\end{equation}
This is a model of a topological 1-dimensional superconductor that supports majorana modes at edges and interfaces, known to be useful for topological computation \cite{Sarma2015MajoranaZM}. In \eqref{Eq:HamKitaev}, $\Delta=\Delta_x + \imath \Delta_y$ is the complex superconducting order parameter, $\mu$ is the on-site chemical potential, $t$ is a hopping parameter, and $S$ is the shift operator on the 1-dimensional lattice while $\sigma$'s are Pauli's matrices. In the bulk, the model displays a spectral gap if $|\mu| \neq |t|$ and a topological phase if $|\mu| < |t|$.

Hamiltonian \eqref{Eq:KitaevReal} is an algebraically equivalent transformation of \eqref{Eq:HamKitaev} which involves only real-valued couplings, yet \eqref{Eq:KitaevReal} preserves all original fundamental symmetries, that is, particle-hole, time-reversal and chiral symmetries. This enables us to implement the fully general Kitaev model with passive metamaterials, in particular, with acoustic resonators (see \cite{PhysRevResearch.7.023143} for an ingenious theoretical mechanical implementation). Although 1-dimensional, the topological interface modes of the chain can be braided using the T-junction technique \cite{PhysRevX.6.031016}. It was pointed in \cite{emilbraiding} that this technique requires a rotation of $\Delta$ in the complex plane in order to keep the spectral gap open while closing the adiabatic cycle of the braiding. Consequently, $\Delta$ cannot be fixed to a real value if we want to achieve the braiding. 

We point out that significant simplifications occur if $\Delta$ is assumed real-valued. Nevertheless, in such simplified setting, \cite{spinnerKitaev} demonstrated experimentally an almost perfect control over the phases and amplitudes of majorana-like modes bound to the edges of a mechanical spinner chain.  Similar experimental control has been demonstrated for acoustic chains \cite{kitaev1} and optical ring-resonator chains \cite{PhysRevResearch.3.013122}. However, the fully general Kitaev Hamiltonian with complex order parameter $\Delta$ has never been implemented experimentally. 

In this work, we fill this gap using acoustic resonators. Our first goal is to demonstrate the fidelity of the experimental platform when it comes to reproducing the symmetries and the phase diagram of the model. Our second goal is to demonstrate the emergence of Majorana-like topological modes at abrupt as well as smooth interfaces. This is important because only the latter can be shifted along the chain without generating non-adiabatic effects. In fact, our third goal is to demonstrate that we can transport the Mojorana-like modes smoothly and subject them to complex rotations of the order parameter without breaking the fundamental symmetries or introducing non-adiabatic effects. 

The required laboratory model implementing Kitaev's chain with complex-valued $\Delta$ is extremely complex, since each unit cell contains four resonators and each resonator must be fitted with five connections (see figure~\ref{fig:h-res-kitaev.svg}). Additionally, the smooth domain-walls require variable geometries from site to site, adding to the complexity. While in a coupled-mode theoretic framework it is always possible to translate tight-binding Hamiltonians to laboratory models, there are assumptions which need to be validated experimentally, such as non-interference of the five mentioned resonator connections, fidelity of the correspondence between the theoretical coupling strengths and geometry of physical connections, or absence of non-adiabatic effects. With the validations supplied by our work in the stated context, the implementation of the full Majorana-braiding cycle reduces to scaling-up the fabrication process, as explained in our discussion section.

\section{Results}

We follow \cite{PhysRevB.98.094310} and amplify $\imath = \sqrt{-1}$ to the real-valued matrix $\imath \sigma_2$ and $1$ to the $2\times 2$ identity matrix. After observing that $\sigma_2$ in \eqref{Eq:HamKitaev} can be written as $-\imath (\imath \sigma_2)$ with the matrix in the parenthesis being real-valued, the original Hamiltonian becomes
\begin{equation}
\label{Eq:KitaevReal}
\begin{aligned}
 & \tilde H= \tfrac{\imath}{2}\sigma_2 \otimes ( \Delta_{x} \sigma_1+ \Delta_y \sigma_3) \otimes (S - S^\dagger) \\
 & \qquad \qquad \qquad \quad - \sigma_2 \otimes \sigma_2 \otimes \big (\mu - \tfrac{t}{2}(S + S^\dagger) \big ).
 \end{aligned}
\end{equation}
It has only real-valued couplings now, but this is achieved at the expense of doubling the degrees of freedom (hence a jump from 2 to 4 modes per repeating cell). As shown in \cite{PhysRevB.98.094310}, after these changes, the energy spectrum remains the same in the bulk and in the presence of interfaces, but its degeneracy doubles. However, the amplified model comes with a built-in symmetry whose symmetry sectors split and decouple those double degeneracies. Furthermore, if $\Pi_\pm$ are the projections into those symmetry sectors, then $\Pi_+ \tilde H \Pi_+$ is unitarily equivalent with $H$ and, as such, it retains all fundamental symmetries \cite{emilbraiding}. The advantage of using $\tilde H$ is that it can be implemented with passive metamaterials, as explained next.  

\subsection{Kitaev Chains with Complex-Valued $\Delta$}

\begin{figure*}[!t]
   \centering
   \def\svgwidth{\textwidth}
   {\fighel
       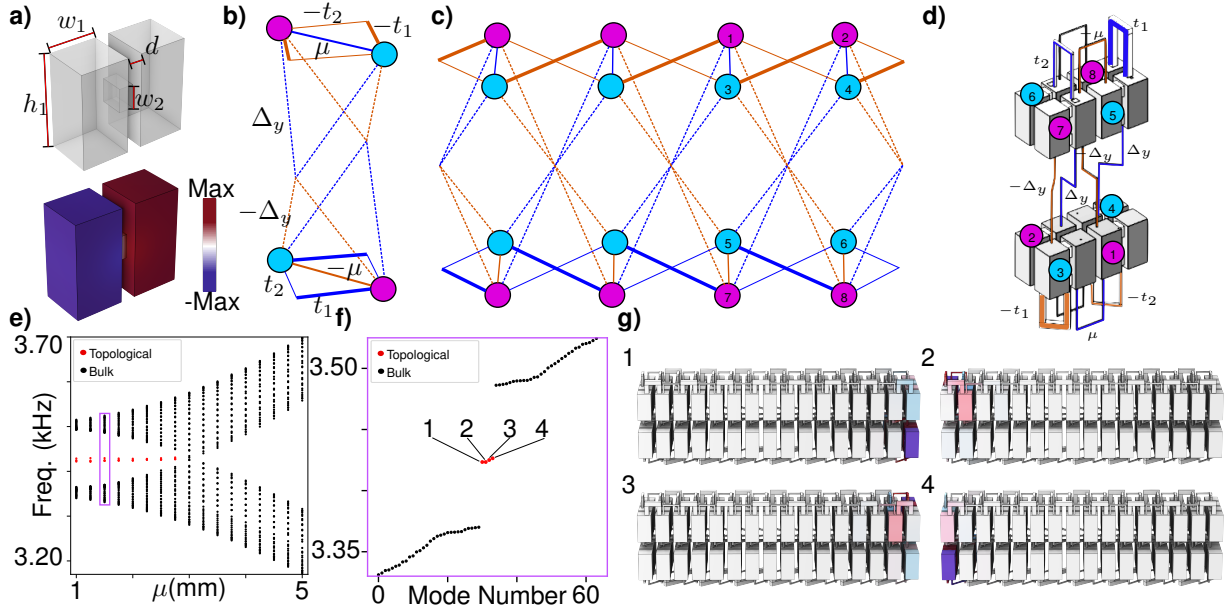
        }
    \caption{\textbf{Acoustic Kitaev chain with complex-valued $\Delta$: Schematic and Simulations}. 
    \textbf{(a)} Geometry of H-resonators: $h_1=20$ mm, $w_1 =$ 10 mm, $d$ = 3 mm, and $w_2=$ 5 mm, and simulated acoustic pressure profile at resonant frequency 3.38 kHz. 
    \textbf{(b)} Single unit cell schematic of the amplified Kitaev chain \eqref{Eq:KitaevReal}, with $t_1=t+\Delta_x$ and $t_2=t-\Delta_x$. Blue (orange) connections represent positive (negative) couplings.
    \textbf{(c)} Full schematic of the amplified Kitaev chain \eqref{Eq:KitaevReal}, with the color coding identical to that in (b). The last 8 resonators are labeled in order to specify their positions in the actual laboratory model shown in (d).
    \textbf{(d)} Section of our laboratory model showing a pair of adjacent acoustic unit cells, with the couplings and resonators  labeled according to the schematics (b-c) (some features exaggerated for clarity).
    \textbf{(e)} Simulated resonant spectrum of the laboratory model as function of the $\mu$-coupling, revealing a topological phase transition {\it exactly} at $\mu = t$, as in the theoretical Kitaev chain \eqref{Eq:HamKitaev}. The couplings $t$, $\Delta_x$, and $\Delta_y$ were fixed at 3 mm, 1 mm, and 1 mm, respectively. Topological edge modes are colored in red and bulk modes in black.
    \textbf{(f)} Resonant spectrum for the $\mu$-coupling value selected in panel (e), with four mid-gap modes labeled and highlighted in red.
    \textbf{(g)} Pressure fields of the mid-gap modes from panel (f), confirming that they are topological edge modes.
    \textbf{(b-c)} are reproductions from \cite{emilbraiding}.}
    \label{fig:h-res-kitaev.svg}
\end{figure*}

To realize the fully general Kitaev model, we use a phononic crystal platform. Our building blocks are the H-resonators seen in figure \ref{fig:h-res-kitaev.svg}(a), composed of two rectangular cavities joined by a single bridge. This allows for the implementation of positive and negative couplings as dictated by the Hamiltonian \eqref{Eq:KitaevReal}. We designed two laboratory models: one based on monolithic 3D printed pieces with sculpted cavities and coupling channels, and a second one based on individually printed H-resonators coupled via silicone tubing. The first one enables extremely precise implementation of site specific couplings, while the second one provides flexibility where needed. By using both of these designs, it is experimentally possible to implement multilayer systems with a dense network of connections at a small scale, as dictated by Hamiltonian \ref{Eq:KitaevReal}. The strengths of the couplings are determined by the widths of the connecting channels or silicon tubes and, as such, they will be quantified in millimeters (mm) throughout. The transfer function connecting the theoretical coupling strengths with the widths of the physical connections has been carefully calibrated before the simulations and experiments started.

Schematic of the unit cell of the amplified model is shown in figure \ref{fig:h-res-kitaev.svg}(b) and the full array of couplings are displayed in figure \ref{fig:h-res-kitaev.svg}(c), as reproduced from \cite{emilbraiding} and derived from Hamiltonian \eqref{Eq:KitaevReal}. Figure \ref{fig:h-res-kitaev.svg}(d) provides an example of two coupled unit cells in the actual acoustic crystals. Note that the resonators are re-positioned in the actual acoustic crystal to simplify the web of couplings.

\begin{figure*}[!t]
   \centering
   \def\svgwidth{\textwidth}
   {\fighel
        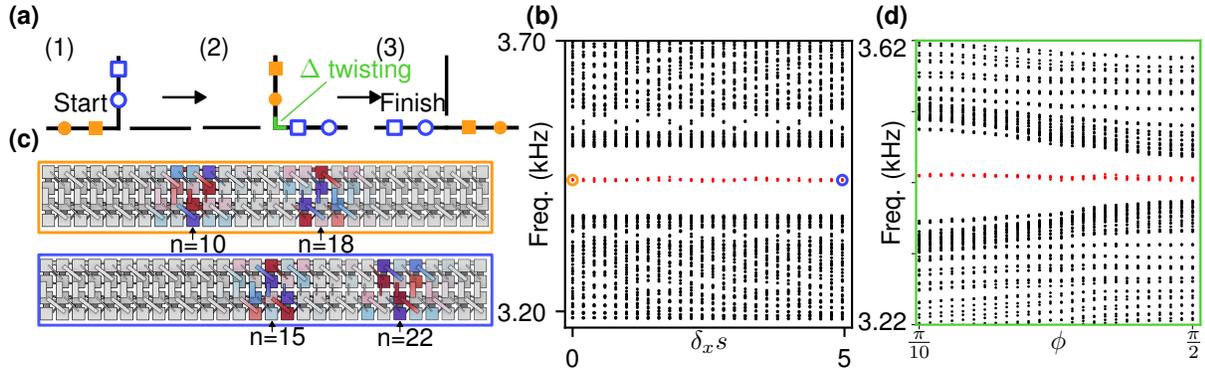
        }
    \caption{
    {\bf Braiding with T-junction Geometry and Validation Simulations.} \textbf{(a)} Braiding process of topological interface modes depicting the three steps described in text. In step (2) of the braid, a twisting of the $\Delta$ parameter is required. \textbf{(b)} Simulated resonant spectrum of a laboratory model partially implementing step (1) using the site-dependent $\mu$-profile \eqref{eq:mu}. The two domain walls were simultaneously shifted by a $\delta_x$ in steps of 0.2, while holding constant $\phi$, $\mu_{\text{min}}$, $\xi$, $x_1$, $x_2$, and $t$ at $\pi/4$, 1 mm, 2 mm, 10, 18, and 2 mm respectively. Topological (bulk) modes are shown in red (black). \textbf{(c)} Pressure fields of the mid-gap resonances highlighted in (b) (note the color coding), corresponding to the initial and final configurations, confirming that the pair of interface modes have been displaced as desired. \textbf{(d)} Simulated resonant spectrum of the laboratory model while twisting the order parameter $\Delta=\Delta_0(\cos \phi +\imath \sin \phi)$ with a variable $\phi$, an operation needed at step (2) of the braiding.}
    \label{fig:braiding-schematic.svg}
\end{figure*}

To demonstrate the fidelity of our laboratory model, we simulate the acoustic crystal with 28 unit cells and for a range of $\mu$-coupling values, and the results reported in figure \ref{fig:h-res-kitaev.svg}(e-g). The full simulated laboratory model can be seen in figure \ref{fig:h-res-kitaev.svg}(g). Figure \ref{fig:h-res-kitaev.svg}(e) displays the simulated resonant spectrum as a function of $\mu$, reproducing the phase diagram of the theoretical Kitaev chain. Indeed, a phase transition can be observed exactly at $\mu =t$ and, for $\mu<t$, mid-gap modes emerge while the spectral gap remains clean for $t > \mu$. Furthermore, the spatial profiles of the mid-gap modes reported in figure \ref{fig:h-res-kitaev.svg}(g) confirm that these modes are localized at the edges of the chain. This confirms without doubt that phase transition seen in figure \ref{fig:h-res-kitaev.svg}(e) is topological. Furthermore, the particle-hole symmetry of the original model is also present here, as evidenced by the high-symmetry of the seen spectra relative to the mid-gap line.

\subsection{Elements of Braiding}
Topological braiding refers to the process of weaving the world lines of non-abelian anyons to form the fundamental logic gates for topological quantum computation. For 1-dimensional systems, braiding can be implemented using the T-junction geometry \cite{PhysRevX.6.031016} illustrated in figure \ref{fig:braiding-schematic.svg}(a). The process consists of displacing two interfaces that trap the topological modes, while coupling and decoupling three strands of topological chains. Specifically, there are three steps required for braiding. (1) The states are prepared on chain 1 and then translated to chain 2 after chain 1 and 2 are fused. (2) Chain 1 and 2 are decoupled and chain 2 and 3 are fused, and the states are migrated from chain 2 to chain 3. Note that chain 3 has an inverted configuration since it has been connected at the ``wrong" end (this is inescapable). Inversion against the middle of a chain is equivalent to changing the sign of $\Delta$. Thus, as explained in \cite{emilbraiding}, we can fix the orientation of chain 3 by varying $\Delta$ until its sign changes. To keep the bulk gap open, this needs to be done in the complex plane. (3) Chain 2 and 3 are decoupled and chain 3 and 1 are fused, and the states are translated back to chain 1 having been swapped. Notice that this procedure has 2 key steps: translation and the $\Delta$ twist. For braiding to work, both of these steps must ensure the mid-gap modes remain spectrally separated and that no non-adiabatic effects are introduced throughout the procedure, which we demonstrate below. 

\begin{figure*}[!t]
  \centering
  \def\svgwidth{\textwidth}
  {\fighel 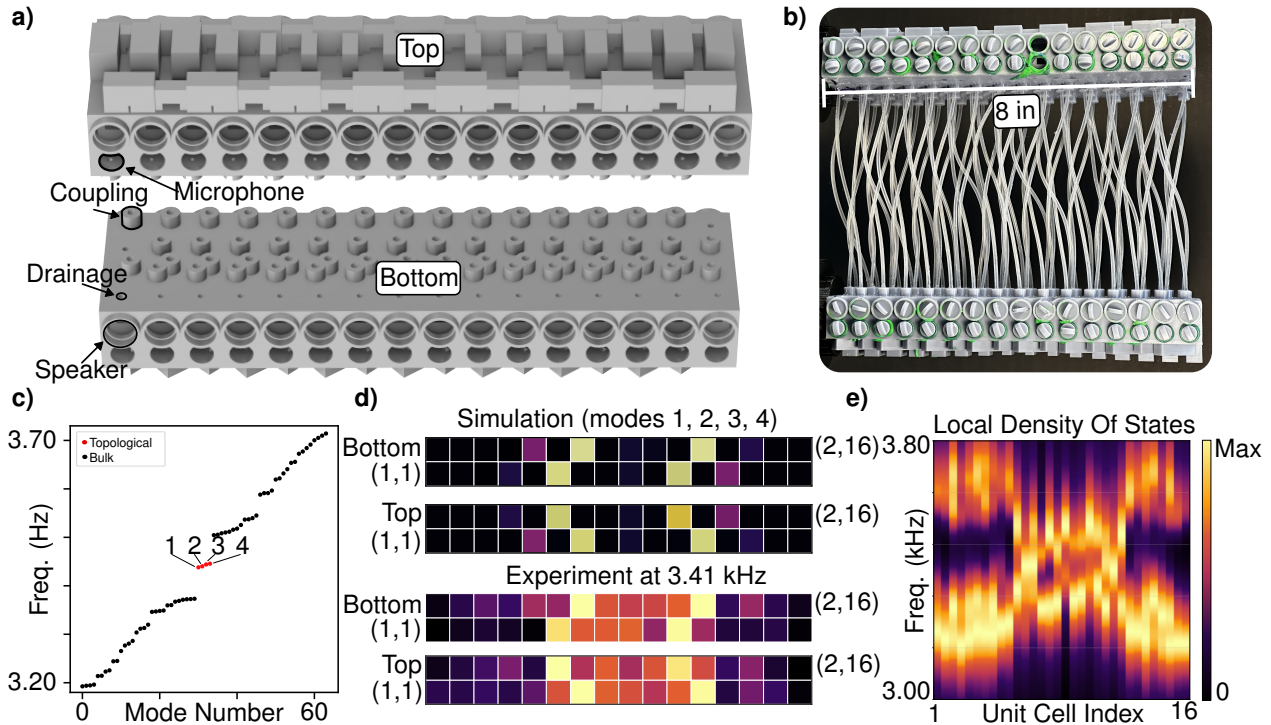}
  \caption{\textbf{Experimental Kitaev System with  local density of states measurements}. \textbf{(a)} STL of the experimental setup showing the two-layer acoustic resonator arrays uncoupled. Each resonator has ports for a microphone, drainage, and a speaker. Coupling ports are placed as dictated by the Hamiltonian. \textbf{(b)} Photograph of the fabricated system with flexible tubing connecting the two layers. \textbf{(c)} Simulated eigenfrequencies showing clear gap with topological modes (red) centered between bulk modes (black). Four topological edge modes are labeled 1-4. \textbf{(d)} Spatial amplitude distribution of the topological edge mode at 3.41 kHz comparing numerical simulations (top panels for each layer) with experimental measurements (bottom panels for each layer). The peak amplitude in the gap is localized at the domain wall. \textbf{(e)} Local density of states (LDOS) measured across all 64 resonators as a function of frequency, confirming spectral gaps closings and openings at the location of the domain walls.}
  \label{fig:majorana-exp}
\end{figure*}

To ensure a smooth deformation of the Hamiltonian while shifting the domain walls, we consider as in \cite{emilbraiding} the following smooth theoretical site-dependence for $\mu$:
\begin{equation}
    \label{eq:mu}
   \mu(n) = \mu_{\text{min}} + 2\xi + \xi\sum_{i=1,2}(-1)^i\tanh\left(\tfrac{n-L-x_i}{l}\right)
\end{equation}
where $\mu_{\text{min}}$ is the minimum physical value of $\mu$-coupling, which for practical reasons cannot be set to 0, $\xi$ is a scaling factor which translates the theoretical values to experimentally viable parameters, $n$ is the index of the unit cell of the lattice, $x_1$ and $x_2$ are the centers of the domain walls, $L$ is the combined number of unit cells of chain 1 and chain 2, and $l$ controls the sharpness of the interfaces and is fixed at 1.

To demonstrate the feasibility of leg (1) of the braiding process, we use Eq.~\eqref{eq:mu} to adjust the parameters of the laboratory model in order to create interfaces. We shift the centers of these interfaces simultaneously, using the rule $x_i = x_i^0 +\delta_x$, where $\delta_x$ is a parameter that will be varied pseudo-continuously. The required geometries of the physical couplings were derived and implemented using the COMSOL and MATLAB live link for maximum parametric flexibility \cite{COMSOL, MATLAB}. The simulated resonant spectrum of the laboratory model as function of $\delta_x$ is reported in figure \ref{fig:braiding-schematic.svg}(b). As seen in the figure, the particle hole-symmetry of the spectrum is retained throughout and the mid-gap modes remain centered nicely in the gap with the degeneracy virtually un-split. Furthermore, the seen spectral features vary smoothly with $\delta_x$, assuring the adiabatic character of the process. The simulated pressure field reported in the orange box in figure \ref{fig:braiding-schematic.svg}(c) confirms that, for $\delta_x=0$, our laboratory model traps mid-gap modes at exactly $x_1^0$ \& $x_2^0$, and transports these modes across the chain as evidenced by the pressure field reported in the blue box in figure \ref{fig:braiding-schematic.svg}(c). 

To validate the fidelity of the laboratory model under complex $\Delta$ twistings, we parametrize it as $\Delta =\Delta_0(\cos{\phi} + i\sin{\phi})$ and let $\Delta_x = \Delta_0\cos{\phi}$ and $\Delta_y = \Delta_0\sin{\phi}$. We sweep $\phi$ from $\pi/10$ to $\pi/2$ while holding all the other parameters at the values specified in figure \ref{fig:braiding-schematic.svg}. This covers a large swath of complex values for $\Delta$. Taking $\phi \to 0$ would cause the $\Delta_y$ couplings to approach 0 width causing meshing errors with discontinuity at $\phi = 0$. At $\phi = \pi/2$, $\text{Re}(\Delta) = 0$, and $t_1$ \& $t_2$ = t. The simulated resonant spectrum of our laboratory model as a function of $\phi$ is reported in figure \ref{fig:braiding-schematic.svg}(d) and this data once again reveals a high degree of symmetry of the spectrum against the mid-gap line and that the topological states remain centered in the gap with degeneracy virtually un-split. Furthermore, the spectral features seen in \ref{fig:braiding-schematic.svg}(d) vary smoothly with the $\Delta$-twist, assuring us of the adiabatic character of the process. 

These simulations validate the designs of our laboratory models as an emulator of the theoretical Kitaev chain and the phenomenologies associated to it. If manufactured with high fidelity, the laboratory model will preserve the fundamental symmetries and the degeneracy of the topological modes as a result, as well as will shield against non-adiabatic effects during continuous shiftings of the domain walls and twistings of the order parameter $\Delta$. In the next section, we demonstrate that is indeed within reach.

\subsection{Experimental Validation}

To experimentally validate the laboratory model, it is enough to manufacture one generic configuration containing two domain walls generated by varying the $\mu$ couplings according to equation \eqref{eq:mu} while fixing $\Delta$ to a complex value. Given size limitations imposed by our 3D printers, we scale the system down from 28 to 16 unit cells with domain walls placed at $x_1 = 5$ and $x_2 = 11$. The fabricated system is shown in Figure \ref{fig:majorana-exp} (a-b), consisting of two layers of 32 H-resonators each, connected via flexible acrylic tubing to implement the $\Delta_y$ couplings between layers (see figure~\ref{fig:h-res-kitaev.svg}). As for the other parameters, we use the same geometry for the H-resonator as in figure \ref{fig:h-res-kitaev.svg}(a), and $\phi$, $l$ $\mu_{\text{min}}$, $\xi$, $x_1$, $x_2$, and $t$ are held constant at $\pi/4$, 1, 1.5 mm, 2 mm, 10, 18, 2 mm respectively.

Each resonator was measured individually using the protocol described in Methods, yielding frequency-amplitude data from 3.00 kHz to 4.00 kHz encoding the local density of states of the wave operator. As a reference, we simulated the resonant spectrum of the specified configuration, which is reported in figure~\ref{fig:majorana-exp}(c), showing the expected four topological mid-gap modes (labeled 1-4 in red) clearly separated from the bulk modes (black), centered around 3.41 kHz within a spectral gap spanning approximately 3.20-3.70 kHz. Given this information, we have measured the local density of states at the frequency of 3.41~kHz, and the result is compared in figure~\ref{fig:majorana-exp}(d) with the overlap of the simulated pressure fields of the four mid-gap modes seen in figure~\ref{fig:majorana-exp}(c). The experimental data shows excellent agreement with the simulations, confirming that the modes are indeed localized at the two domain walls (resonators 5 and 11) with characteristic decay into the bulk regions. The localization length is approximately one lattice site in both directions, matching theoretical predictions for Majorana-like edge modes.

The full local density of states (LDOS) across all 32 resonators in the top layer, shown in Figure \ref{fig:majorana-exp} (e), reveals the \st{complete} spectral landscape of the system. By construction, the domain wall forces the system to enter a sequence of trivial-topological-trivial phases, with spatial transitions at the domain walls. The seen experimental LDOS captures all of that, {\it e.g.} the opening of the bulk spectral gaps away from the transitions is clearly visible, and same for the topological interface modes trapped at the transitions. The spectral gap of the topological phase is smaller than the one of the trivial phase, which is a consequence of the asymmetry of the phase diagram reported in figure~\ref{fig:h-res-kitaev.svg}(e) as a function of $\mu$-coupling.

These experimental results confirm that our laboratory model can be fabricated with enough fidelity to display the key expected features, notably, the openings of the bulk spectral gaps and the emergence of the topological interface modes under a modulation of the pseudo chemical potential. Future iterations of the fabrication process will improve the fidelity, hopefully to a point where we will see a better spatial uniformity away from the interfaces and a clearer particle-hole symmetry of the spectrum.

\section{Discussions}
In this work, we developed a laboratory model based on an acoustic resonator platform for the fully general theoretical Kitaev chain model with complex order parameter. The simulations with the acoustics module of COMSOL Multiphysics \cite{COMSOL} indicate that the transfer from the theoretical model to our laboratory model achieves an extremely high fidelity under a series of tests. These tests were tailored to the task of braiding the Majorana-like interface modes, which we are now convinced that it can be achieved with our laboratory model. Furthermore, we have given evidence that the laboratory model can be already fabricated with enough precision to reproduce many of the expected key features. We are convinced that future iterations of the fabrication process will bring the simulations and the experiment to equal footing.

As for the full brading cycle of the Majorana-like interface modes, it is now just a matter of scaling up the simulations and the fabrication process. Indeed, the braiding cycle can be completed dynamically in time, in which case one needs a reconfigurable laboratory model, but the cycle can be also rendered in space \cite{ChenNC2021}. In fact, this was already demonstrated in \cite{Chen2022} in a continuous model setting. In our discrete case, an array of weakly-coupled laboratory models, with different parameters set by the braiding cycle, will implement the adiabatic cycle, according to the WKB-type analysis from \cite{WKBAnalysis}. The results of our present work assures us that each layer in the mentioned array of laboratory models will perform exactly as expected, thus achieving braiding of Majorana-like topological modes is indeed a matter of scaling up the fabrication process.

\section{Methods}
\subsection{Fabrication}
The fabrication of the system is done using resin 3D printers with UV curable resin. The top and bottom layers of the system are printed individually. The drain holes, inserted to prevent resin buildup in the closed chambers, are then plugged using modeling clay. Flexible acrylic tubing is then glued into 3D printed ports to connect the two halves. Plugs are printed to close the microphone and speaker ports when not in use. This method allows for a dense packing of couplings and resonators allowing for easy scalability. 

\subsection{Experimental Protocol}
Each resonator is measured by placing a microphone and speaker in a single chamber of the resonator. The speaker then excites a sine wave with a constant amplitude at frequencies ranging from 3.00 kHz to 4.00 kHz in steps of 0.01 kHz. These signals are generated using a custom Python script with the output coming directly from the 1.5 mm jack in the computer. The same script uses an Audix TM1 Plus microphone that is connected via USB by an M-track Audio Solo to record the signal at each frequency and calculates the root mean square of the measured pressure values giving us a single amplitude for a given frequency. This is then stored as a CSV file on the computer.

\subsection{Simulations}
All simulations were conducted using finite element analysis in the acoustics module of COMSOL Multiphysics\cite{COMSOL}. The pressure profiles were filled with air with a density of 1.3 kg/m\textsuperscript{3} and speed of sound at 343 m/s, appropriate for room temperature. Given the large acoustic impedance mismatch between the air and the UV Acrylic, the boundaries were treated as hard walls. Throughout this paper, loss was not modeled.

\section{Acknowledgments}
C.P. and J.S. acknowledge support from the U.S. National Science Foundation through the grant CMMI-2414984, and by U.S. Army Research Office through contract W911NF‐25-1-0050. E.P. acknowledges support from the U.S. National Science Foundation through the grant CMMI-2131760, and by U.S. Army Research Office through contract W911NF-23-1-0127.


\bibliographystyle{naturemag}
\bibliography{refrences}

\clearpage
\appendix
\renewcommand{\thesection}{S\arabic{section}}
\renewcommand{\thesubsection}{S\arabic{section}.\arabic{subsection}}
\renewcommand{\theequation}{S\arabic{equation}}
\renewcommand{\thefigure}{S\arabic{figure}}
\renewcommand{\thetable}{S\arabic{table}}

\setcounter{section}{0}
\setcounter{equation}{0}
\setcounter{figure}{0}
\setcounter{table}{0}





\end{document}